\documentclass{aa}
\usepackage[usenames,dvipsnames]{xcolor}
\usepackage{natbib}
\usepackage{xspace}
\usepackage{amssymb}
\usepackage{amsmath}
\usepackage{graphicx}
\usepackage{hyperref}

\def\comma{\,,}
\def\fullstop{\,.}
\newcommand{\revised}[1]{{#1}}
\newcommand{\revisedtwo}[1]{{#1}}
\begin{document}
\title{Dust-driven viscous ring-instability in protoplanetary disks}  \titlerunning{Viscous instability in protoplanetary disks}
\authorrunning{Dullemond and Penzlin} 
\author{C.P.~Dullemond and A.B.T.~Penzlin}
\institute{Zentrum f\"ur Astronomie, Heidelberg University, Albert Ueberle Str.~2, 69120 Heidelberg, Germany} \date{\today}

\abstract{Protoplanetary disks often appear as multiple concentric rings in dust
  continuum emission maps and scattered light images. These features are often
  associated with possible young planets in these disks.  Many non-planetary
  explanations have also been suggested, including snow lines, dead zones and
  secular gravitational instabilities in the dust. In this paper we suggest
  another potential origin. The presence of copious amounts of dust tends to
  strongly reduce the conductivity of the gas, thereby inhibiting the
  magneto-rotational instability, and thus reducing the turbulence in the disk.
  From viscous disk theory it is known that a disk tends to increase its surface
  density in regions where the viscosity (i.e.~turbulence) is low. Local maxima
  in the gas pressure tend to attract dust through radial drift, increasing the
  dust content even more. We have investigated mathematically if this could
  potentially lead to a feedback loop in which a perturbation in the dust
  surface density could perturb the gas surface density, leading to increased
  dust drift and thus amplification of the dust perturbation and, as a
  consequence, the gas perturbation. We find that this is indeed possible, even
  for moderately small dust grain sizes, which drift less efficiently, but which
  are more likely to affect the gas ionization degree. We speculate that this
  instability could be triggered by the small dust population initially, and
  when the local pressure maxima are strong enough, the larger dust grains get
  trapped and lead to the familiar ring-like shapes. We also discuss
  the many uncertainties and limitations of this model.
}

\maketitle


\section{Introduction}
High resolution, high contrast imaging at optical and sub-millimeter wavelengths
has in recent years revealed that protoplanetary disks are highly structured.
While some disks show complex non-axisymmetric structures, there are also
numerous disks that display multiple nearly perfect concentric ringlike
structures. The first and most prominent example was the disk around the star HL
Tau \citep{2015ApJ...808L...3A}. Many other examples have since followed such as
TW Hydra \citep{2016ApJ...820L..40A,2017ApJ...837..132V}, RX J1615.3
\citep{2016A&A...595A.114D}, HD 97048 \citep{2016A&A...595A.112G}, HD 163296
\citep{2016PhRvL.117y1101I} and HD 169142 \citep{2015PASJ...67...83M,
  2017A&A...600A..72F}. The rings are seen at submillimeter wavelength thermal
dust emission as well as in optical/near-infrared scattered light. These rings
and gaps have been interpreted as resulting from newly formed planets opening up
gaps within the disk \citep[e.g.][]{2015MNRAS.454L..36G, 2015ApJ...806L..15K,
  2015A&A...584A.110P}. This is an attractive scenario, because it would mean
that we are indirectly `seeing' the young planets as they are being formed in
their birth-disk. However, since we do not yet have direct indications of these
planets, it is important to also investigate other explanations. Could these
rings be caused by something entirely different altogether? And could this
`something', though unrelated to already existing planets, still teach us
something about the formation of planets?

Various non-planet-related explanations for these rings have been proposed.  In
fact, their existence was predicted before their discovery, through a simple
argument: It is known that dust aggregates of millimeter size and larger tend to
radially drift toward the star on a very short time scale
\citep{1972fpp..conf..211W, 2007A&A...469.1169B}. If the disk lives a few
millions years, and the dust aggregates grow to the millimeter sizes inferred
from millimeter-wave observations \citep[e.g.][]{2016A&A...588A..53T}, then
these disks should by the time they are observed already have lost most of their
dust particles. By contrast, we see large quantities of dust in these disks, so
something must be holding up the dust, preventing it from taking part in the
rapid radial drift mechanism. It was suggested by \citet{2012A&A...538A.114P}
that perhaps a multitude of local ring-shaped gas pressure bumps in the disk
could trap the dust after it has grown to sizes of about a millimeter. The
physical mechanism behind this dust trapping is the well-known, and unavoidable
effect, that dust particles tend to drift toward regions of increased gas
pressure \citep[e.g.][]{1972fpp..conf..211W, 1976PThPh..56.1756A}. Or in other
words, that they drift in the direction of the gas pressure gradient $\vec\nabla
P$. In smooth protoplanetary disks the gas pressure decreases with radius
($\partial_r P<0$), which is the reason for the inward drift of dust. But if the
gas pressure has wiggles that are strong enough that they cause $\partial_r P$
to flip sign, then there will exist local pressure maxima for which $\partial_r P=0$,
and for which the dust drift is converging. In the absense of gas turbulence,
all dust grains sufficiently close by would get trapped in these
traps. Turbulence can mix the small dust grains (that are most strongly coupled
to the gas) out of the traps, so that they may, on average, continue to drift
inward. The bigger dust particles, which are less coupled to the gas, may still
remain trapped. It depends on the amplitude of the pressure bumps and the
strength of the turbulence, which grain sizes remain trapped and which not.

While the origin of these gas pressure bumps was not addressed in
\citet{2012A&A...538A.114P}, it was shown that if they exist, and if they
are strong enough, the radial drift paradoxon could be solved. As a direct
consequence, however, it was shown that high-resolution ALMA observations should
then be able to see these dust rings, and the predicted ALMA images bear
striking resemblance to HL Tau and similar sources.

This scenario does not, however, explain why these gas pressure bumps form
in the first place. \cite{2014ApJ...794...55T, 2016AJ....152..184T} propose an
elegant scenario in which dust rings are formed through a secular gravitational
instability \citep{2000orem.book...75W}. The idea is that if the dust density is
high enough for a dust-driven gravitational instability to occur, the gas drag
will slow this process down. The slowness of this process allows the information
about the gravitational contraction to shear out and spread along azimuth, so
that grand-design rings are formed instead of gravitationally contracting
clumps.  \citet{2016AJ....152..184T} argue that as these rings contract further,
this eventually leads to planets being formed, which, in their turn, open up
gaps in the dust distribution \citep{2004A&A...425L...9P}. They suggest that the
ring-like structures could therefore be witnesses of both the initial and the
final stages of planet formation.

A completely different scenario was proposed by \citet{2015ApJ...806L...7Z}, who
argue that the locations of the rings suggest their association with the snow
lines of a series of different volatile molecular species. A physical mechanism
by which snow lines could lead to rings was worked out by
\citet{2016ApJ...821...82O}. The physics of ice sublimation and deposition near
these snow lines is complex, because it interacts strongly with the coagulation
and fragmentation of dust aggregates, as well as with the radial drift and
turbulent mixing in the disk \citep[e.g.][]{2017A&A...600A.140S}.

\citet{2017MNRAS.467.1984G} propose an alternative scenario of spontaneous ring
formation. In their model the radial drift of dust particles coupled to the dust
coagulation process tends to lead to ring-like regions of high dust
concentration and fast growth \citep[see also][]{2016A&A...594A.105D}.  While
this might explain transition disks with a single dust ring, it may be more
difficult to explain the multi-ring structures discussed here.

Global magnetohydrodynamical disk simulations with dead-zones also tend to
create rings-shaped structures \citep{2015A&A...574A..68F}, and zonal flows
\citep{2009ApJ...697.1269J}.

In this paper we investigate whether the viscous disk evolution could lead to
the spontaneous formation of rings. This idea is not new. For instance,
\citet{2005MNRAS.362..361W} show, using 2-D radiation-hydrodynamics models, that
a disk with an active surface layer, self-gravity and `dead' midplane region
could become viscously unstable and lead to the formation of several concentric
rings. Our proposal is, however, based on a different physical driving
mechanism. A version of this mechanism was already studied in a local box model
by \citet{2011IAUS..274...50J}.

\begin{figure*}
  \centerline{\includegraphics[width=0.94\textwidth]{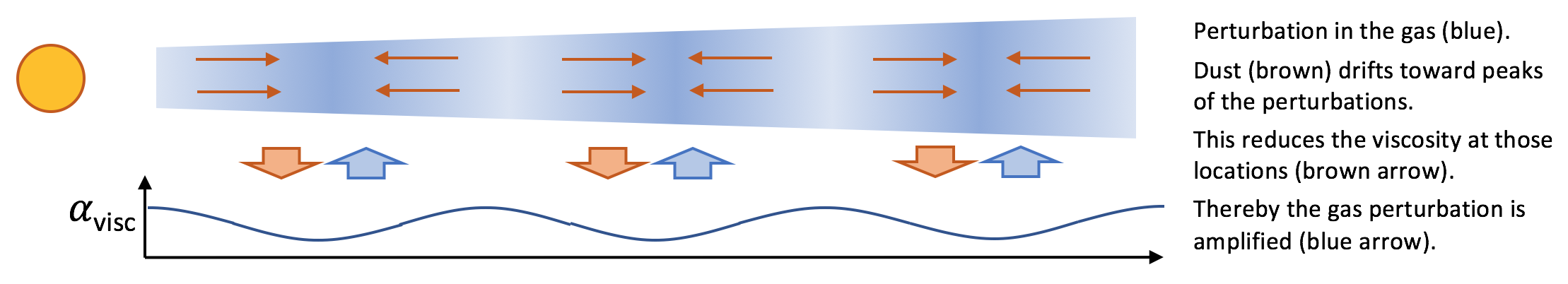}}
  \caption{\label{fig-cartoon}Mechanism of the ring
    instability studied in this paper.}
\end{figure*}

The ring instability works as follows. If we perturb an otherwise
smooth disk with an infinitesimal-amplitude wiggle in the gas pressure (of the
concentric ring type), it will tend to cause a slight enhancement of the
dust-to-gas ratio in these pressure enhancements. This is the same physical
mechanism as for dust trapping, but we do not necessarily need a flip of sign of
$\partial_r P$ to cause this effect. It is just that the radial inward drift
velocity $|v_{\mathrm{drift}}|$ of the dust is slightly increased on the outer
side of the pressure enhancement and slightly reduced on the inner side, leading
to a traffic-jam density enhancement effect. This effect works for big and small
grains alike. It is known that dust has a negative influence on the disk
viscosity, if it is caused by the magnetorotational instability
\citep[e.g.][]{2000ApJ...543..486S, 2006A&A...445..205I, 2009ApJ...698.1122O,
  2013ApJ...765..114D}. A viscous disk reacts to the resulting wiggle in
the viscosity $\nu(r)$ by adapting the surface density $\Sigma_g(r)$ such that
the steady-state
\begin{equation}
\Sigma_g(r)\,\nu(r) = \mathrm{const}\comma
\end{equation}
is restored. So, whereever $\nu$ is reduced, $\Sigma_g$ is increased. This change 
in $\Sigma_g(r)$ amplifies the initial perturbation, resulting in a positive feedback
loop. We therefore expect an initial perturbation in the gas disk to be amplified
by the combined effect of dust drift and the effect the dust has on the viscosity
of the disk. This process is depicted in cartoon form in Fig.~\ref{fig-cartoon}.

To test whether this mechanism indeed works requires a linear perturbation
analysis. This is what we present in this paper.

In Section \ref{sec-basic-equations} we give the basic equations that stand
at the basis of our model. These are the standard viscous disk equations
coupled to a single dust component of a given Stokes number.

The linear perturbation analysis of the combined gas and dust system is rather
cumbersome. So in Section \ref{sec-pert-analy-simple} we first simplify the
system of equations radically, so that the mechanism presents itself more
clearly. In Section \ref{sec-pert-analy-full} we then tackle the full
set of equations, with some mathematics moved to the appendix.

\section{Basic disk equations and model assumptions}
\label{sec-basic-equations}
The standard viscous disk equations for the gas together with a single dust
component are:
\begin{eqnarray}
\frac{\partial \Sigma_{g}}{\partial t} +
  \frac{1}{r}\frac{\partial}{\partial r}
  \left(r\Sigma_g v_{rg}\right) &=& 0\label{eq-gas-continuity}\\
\frac{\partial \Sigma_{d}}{\partial t} +
  \frac{1}{r}\frac{\partial}{\partial r}
  \left(r\Sigma_d v_{rd}\right) &=&
  \frac{1}{r}\frac{\partial}{\partial r}
  \left(r{\cal D}_d\Sigma_g\frac{\partial}{\partial r}
  \left(\frac{\Sigma_d}{\Sigma_g}\right)\right)\fullstop\label{eq-dust-continuity}
\end{eqnarray}
The gas radial velocity is given by the usual viscous disk equation:
\begin{equation}\label{eq-vr-gas}
  v_{rg} = - \frac{3}{\Sigma_g\sqrt{r}}\frac{\partial}{\partial r}
  \left(\Sigma_g\nu\sqrt{r}\right)\comma
\end{equation}
with the turbulent viscosity defined by
\begin{equation}\label{eq-def-nu}
\nu = \alpha \frac{c_s^2}{\Omega_K}\comma
\end{equation}
where $\alpha$ is the usual alpha-turbulence parameter. The Kepler frequency is
\begin{equation}\label{eq-omega-kepler}
\Omega_K=\sqrt{\frac{GM_{*}}{r^3}}\comma
\end{equation}
with $G$ the gravitational constant and $M_*$ the stellar mass.  The isothermal
sound speed squared is
\begin{equation}\label{eq-cs2-in-temp}
c_s^2 = \frac{k_BT}{\mu m_p}\comma
\end{equation}
with $k_B$ the Boltzmann constant, $\mu=2.3$ the mean molecular weight,
and $m_p$ the proton mass. The vertical pressure scale height of the disk is
\begin{equation}\label{eq-disk-hp}
H_p = \frac{c_s}{\Omega_K}\fullstop
\end{equation}
The dust diffusion constant is
\begin{equation}\label{eq-cald-in-nu}
  {\cal D}_d = \frac{1}{1+\mathrm{St}^2}{\cal D}_g =
  \frac{1}{1+\mathrm{St}^2}\frac{\nu}{\mathrm{Sc}}\comma
\end{equation}
where $\mathrm{St}$ is the Stokes number of the dust and $\mathrm{Sc}$ is the
Schmidt number of the gas. The radial velocity of the dust is
\begin{equation}\label{eq-vdust-1}
  v_{rd} = \frac{1}{1+\mathrm{St}^2} v_{rg}
  + \frac{1}{\mathrm{St}+\mathrm{St}^{-1}}\frac{1}{\rho_g\Omega_K}
  \frac{\partial P_g}{\partial r}\comma
\end{equation}
where $\rho_g$ is the midplane gas density and $P_g\equiv \rho_g c_s^2$ is the
midplane gas pressure.

The $\alpha$ determines the strength of the turbulence, and hence the strength
of the viscosity and the dust diffusivity. In our analysis we allow the dust
surface density to affect the value of $\alpha$: a higher dust concentration
will lead to a lower $\alpha$ \citep[e.g.][]{2000ApJ...543..486S, 2006A&A...445..205I, 2009ApJ...698.1122O,
  2013ApJ...765..114D}. Since the \revised{physics of magnetorotational
  turbulence in non-ideal MHD is not yet fully understood},
we parameterize this effect. We consider the following
general prescription:
\begin{equation}\label{eq-alpha-param}
  \alpha = \alpha_1\left(\frac{\Sigma_d}{\Sigma_{d1}}\right)^{\phi_d}
  \left(\frac{\Sigma_g}{\Sigma_{g1}}\right)^{\phi_g}\comma
\end{equation}
where $\alpha_1$ is the unperturbed value of $\alpha$, and likewise
$\Sigma_{d1}$ and $\Sigma_{g1}$ are the unperturbed values of $\Sigma_d$ and
$\Sigma_g$ respectively.  The parameters $\phi_d$ and $\phi_g$ are powerlaw
indices that parameterize how $\alpha$ depends on the change in dust and/or gas
surface density. We focus on cases with $\phi_d<0$, meaning that an increase
in $\Sigma_d/\Sigma_{d1}$ leads to a {\em de}crease in $\alpha$. The $\phi_g$
is used to allow for the following two cases of interest:
\begin{eqnarray}
\phi_g =& 0 & \qquad \hbox{(Case 1: $\alpha$ depends on $\Sigma_d$)} \label{eq-case-1}\\
\phi_g =& -\phi_d & \qquad \hbox{(Case 2: $\alpha$ depends on $\Sigma_d/\Sigma_g$)} \label{eq-case-2}
\end{eqnarray}
Case 1 could, at least in principle, allow for the expected instability even
without dust drift relative to the gas, since without dust drift, $\Sigma_d$
will necessarily increase if $\Sigma_g$ does. In contrast, case 2 strictly
requires dust drift for the instability to operate, since lack of dust drift
keeps $\Sigma_d/\Sigma_g$ constant.

\section{Simplified perturbation analysis}
\label{sec-pert-analy-simple}
It is cumbersome to perform the linear stability analysis for the full set of
equations of Section \ref{sec-basic-equations}, because these equations contain
factors $\sqrt{r}$ and the like. We postpone this full analysis to Section
\ref{sec-pert-analy-full}. For now, let us first simplify the equations.

\subsection{Simplified equations}
We simplify the equations to the following form:
\begin{eqnarray}
\frac{\partial \Sigma_{g}}{\partial t} +
\frac{1}{r_0}\frac{\partial}{\partial x}
  \left(\Sigma_g v_{xg}\right) &=& 0\comma\label{eq-simpl-gas-continuity}\\
\frac{\partial \Sigma_{d}}{\partial t} +
\frac{1}{r_0}\frac{\partial}{\partial x}
  \left(\Sigma_d v_{xd}\right) &=&
  \frac{1}{r_0^2}
  \frac{\partial}{\partial x}
  \left({\cal D}_d\Sigma_g\frac{\partial}{\partial x}
  \left(\frac{\Sigma_d}{\Sigma_g}\right)\right)\comma\label{eq-simpl-dust-continuity}
\end{eqnarray}
where $r_0$ is the radius at which we wish to locally carry out the perturbation
analysis, and $x$ is the dimensionless radial coordinate starting from that
location:
\begin{equation}
r = r_0 (1+x)\fullstop
\end{equation}
The gas radial velocity formula (Eq.~\ref{eq-vr-gas}) is simplified as:
\begin{equation}\label{eq-simpl-vr-gas}
  v_{xg} = - \frac{3}{\Sigma_gr_0}\frac{\partial(\Sigma_g\nu)}{\partial x}\comma
\end{equation}
with the viscosity still defined by Eq.~(\ref{eq-def-nu}). However, for
simplicity $c_s$ and $\Omega_K$ are now assumed be be constant. On the
other hand, $\alpha$ is allowed to depend on $x$, but only due to the
perturbation. The radial velocity of the dust is given, in our simplified
description, by
\begin{equation}\label{eq-simpl-vdust-1}
  v_{xd} = \frac{1}{1+\mathrm{St}^2} v_{xg}
  + \frac{1}{\mathrm{St}+\mathrm{St}^{-1}}\frac{c_s^2}{\Omega_Kr_0}
  \frac{\partial\ln \Sigma_g}{\partial x}\fullstop
\end{equation}
The stationary solution is $\Sigma_g=$constant, $\Sigma_d=$constant and
$\alpha=$constant. This also yields $v_{xd}=v_{xg}=0$ for that stationary
solution. This stationary solution is the backdrop of our perturbation
analysis. 

Now we introduce an infinitesimal perturbation:
\begin{eqnarray}
\Sigma_g(x,t) &=& \Sigma_{g1} (1+\sigma_g(x,t))\comma\label{eq-simpl-sigma-g-pert}\\
\Sigma_d(x,t) &=& \Sigma_{d1} (1+\sigma_d(x,t))\comma\label{eq-simpl-sigma-d-pert}
\end{eqnarray}
where $\Sigma_{g1}$ and $\Sigma_{d1}$ are the stationary (constant) solutions
for gas and dust, respectively. From here on the subscript $1$ denotes this
stationary solution. We will also omit the $(x,t)$ notation, to not clutter the
equations too much. The symbols $\sigma_g$ and $\sigma_d$ are the dimensionless
infinitesimal perturbations on the gas and the dust respectively. For the
$\alpha$, according to Eq.~(\ref{eq-alpha-param}), we 
obtain, to leading order:
\begin{equation}\label{eq-alpha-prescr}
\alpha = \alpha_1 \big(1+\phi_d\sigma_d+\phi_g\sigma_g\big)\comma
\end{equation}
on account of the fact that $|\sigma_{d/g}|\ll 1$.

Inserting these formulae into Eqs.~(\ref{eq-simpl-gas-continuity},
\ref{eq-simpl-dust-continuity}), and keeping in mind that both $v_{xg}$ and
$v_{xd}$ are linear in the perturbations, and that we omit all terms
of higher order in $\sigma_{g/d}$, we arrive, to leading order, at:
\begin{eqnarray}
\frac{\partial \sigma_{g}}{\partial t} +
  \frac{1}{r_0}\frac{\partial v_{xg}}{\partial x}
   &=& 0\comma\label{eq-simpl-gas-continuity-1}\\
\frac{\partial \sigma_{d}}{\partial t} +
  \frac{1}{r_0}\frac{\partial v_{xd}}{\partial x}
  &=&
  \frac{1}{r_0^2}{\cal D}_d
  \frac{\partial^2(\sigma_d-\sigma_g)}{\partial x^2}\fullstop
  \label{eq-simpl-dust-continuity-1}
\end{eqnarray}
By inserting Eqs.~(\ref{eq-def-nu}, \ref{eq-simpl-sigma-g-pert}) together with
Eq.~(\ref{eq-alpha-prescr}) into Eq.~(\ref{eq-simpl-vr-gas}), and assuming
that $c_s$ and $\Omega_K$ are constant (see above), the radial velocity for the
gas becomes
\begin{equation}
  v_{xg} = - 3\frac{\nu_1}{r_0}\left(\frac{\partial\sigma_g}{\partial x}
  +\frac{\partial(\phi_d\sigma_d+\phi_g\sigma_g)}{\partial x}\right)\fullstop
\end{equation}
Similarly that of the dust becomes
\begin{equation}
\begin{split}
  v_{xd} =&
  \frac{1}{1+\mathrm{St}^2}\frac{\nu_1}{r_0}\bigg\{
  - 3\left(\frac{\partial\sigma_g}{\partial x}
  +\frac{\partial(\phi_d\sigma_d+\phi_g\sigma_g)}{\partial x}\right)\\
&  \qquad\qquad\quad
  +\frac{\mathrm{St}}{\alpha_1}
  \frac{\partial\sigma_g}{\partial x}\bigg\}\fullstop\\
\end{split}
\end{equation}
Inserting these into the continuity equations
(Eqs.~\ref{eq-simpl-gas-continuity-1},
  \ref{eq-simpl-dust-continuity-1}) yields
\begin{eqnarray}
  \frac{\partial\sigma_g}{\partial t}
  &=& 3\frac{\nu_1}{r_0^2}\left(\frac{\partial^2\sigma_g}{\partial x^2}
  +\frac{\partial^2(\phi_d\sigma_d+\phi_g\sigma_g)}{\partial x^2}\right)\comma\\
  \frac{\partial\sigma_d}{\partial t} &=&
  \frac{1}{1+\mathrm{St}^2}\frac{\nu_1}{r_0^2}\bigg\{
  3\left(\frac{\partial^2\sigma_g}{\partial x^2}
  +\frac{\partial^2(\phi_d\sigma_d+\phi_g\sigma_g)}{\partial x^2}\right)
   \nonumber\\
 & &  \qquad\qquad\quad -\frac{\mathrm{St}}{\alpha_1}
  \frac{\partial^2\sigma_g}{\partial x^2}\bigg\}
  + \frac{1}{r_0^2}{\cal D}_{d1}
  \frac{\partial^2(\sigma_d-\sigma_g)}{\partial x^2} \fullstop
\end{eqnarray}
Now let us assume the linear perturbations to be plane waves in space $x$ and
time $t$. Since we seek modes for which the dust and the gas perturbations
growth due to their mutual coupling, we can assume a single spatial frequency
$k$ and time frequency $\omega$ for both modes:
\begin{eqnarray}
\sigma_g &=& A e^{i\omega t-ikx}\comma\\
\sigma_d &=& B e^{i\omega t-ikx}\fullstop
\end{eqnarray}
The complex amplitudes $A$ (for the gas) and $B$ (for the dust) can be
set independently. Inserting this mode into the above set of equations yields:
\begin{eqnarray}
  i\omega A
  &=& -3k^2 \frac{\nu_1}{r_0^2}\left(A
  +\phi_d B+\phi_gA\right)\comma\\
  i\omega B &=&
  \frac{-k^2}{1+\mathrm{St}^2}\frac{\nu_1}{r_0^2}\bigg\{
  3\left(A
  +\phi_dB+\phi_gA\right)
  -\frac{\mathrm{St}}{\alpha_1}A
   \nonumber\\
 & & \qquad\qquad\qquad +\frac{1}{\mathrm{Sc}}
  (B-A)\bigg\}\comma
\end{eqnarray}
\revised{where we made use of Eq.~(\ref{eq-cald-in-nu}) to replace ${\cal D}_{d1}$.}
This can be put into matrix form:
\begin{equation}
i\omega \left(\begin{matrix}
  A\\
  B
\end{matrix}\right)
=\left(\begin{matrix}
  M_{aa} & M_{ab}\\
  M_{ba} & M_{bb}
\end{matrix}\right)
\left(\begin{matrix}
  A\\
  B
\end{matrix}\right)\comma
\end{equation}
with
\begin{eqnarray}
  M_{aa} &=& -3k^2\frac{\nu_1}{r_0^2}(1+\phi_g)\comma \\
  M_{ab} &=& -3k^2\frac{\nu_1}{r_0^2}\phi_d\comma \\
  M_{ba} &=& -3k^2\frac{\nu_1}{r_0^2}
  \frac{1}{1+\mathrm{St}^2}\left((1+\phi_g)-\frac{\mathrm{St}}{3\alpha_1}-\frac{1}{3\mathrm{Sc}}\right)\comma\\
  M_{bb} &=& -3k^2\frac{\nu_1}{r_0^2}
  \frac{1}{1+\mathrm{St}^2}\left(\phi_d+\frac{1}{3\mathrm{Sc}}\right)\fullstop
\end{eqnarray}
The eigenvalues of this matrix are found from:
\begin{equation}\label{eq-gamma-pm}
  \begin{split}
    \Gamma_{\pm} =& \frac{1}{2}\bigg((M_{aa}+M_{bb}) \\
&   \pm \sqrt{(M_{aa}+M_{bb})^2-4(M_{aa}M_{bb}-M_{ba}M_{ab})}\bigg)\fullstop
  \end{split}
\end{equation}
So we have
\begin{equation}\label{eq-iomega-from-gamma}
i\omega = \Gamma_{\pm}\fullstop
\end{equation}
The solution is stable if
\begin{equation}
\mathrm{Re}(i\omega) \le 0\comma
\end{equation}
for all possible \revised{values of $k$. One can see that if this is true/untrue
  for one value of $k$, it is true/untrue for all values of $k$, because $k$
  only enters as a multiplicative factor. The above stability condition
  requires that both $M_{aa}+M_{bb}<0$ and $M_{aa}M_{bb}>M_{ab}M_{ba}$.
  These two conditions simplify to:}
\begin{eqnarray}
  1+\phi_g + \frac{1}{1+\mathrm{St}^2}\left(\phi_d + \frac{1}{3\mathrm{Sc}}\right) &\ge& 0\comma
\label{eq-stability-criterion-simple-1}\\
  1+\phi_g + \left(1 + \frac{\mathrm{St}}{\alpha_1}\,\mathrm{Sc}\right)\phi_d &\ge& 0\fullstop
\label{eq-stability-criterion-simple-2}
\end{eqnarray}
\revised{Both conditions have to be fulfilled for the disk to be stable against
the dust-driven viscous instability. Otherwise it is unstable for all $k$.}

\subsection{Results for the growth rates}\label{sec-results-simple}
We apply the model to the case of a young solar mass star ($M_{*}=M_\odot$) with
still substantial luminosity ($L_{*}=10\,L_\odot$), to mimic the case of HL Tau
\citep{2015ApJ...808L...3A}. We perform the analysis at a radius of
$r_0=60\,\mathrm{au}$. The Schmidt number of the gas is set to $\mathrm{Sc}=1$.
To compute the temperature of the midplane gas we assume a simple irradiated
disk model, in which the irradiation angle is $\varphi=0.05$. By setting the
midplane temperature to the effective temperature of the disk assuming thermal
equilibrium we obtain $T=(\varphi L_{*}/(4\pi r_0^2\sigma_{\mathrm{SB}}))^{1/4}
=43\,\mathrm{K}$, which is a very rough estimate of the disk temperature, but
sufficient for the present purpose. The orbital time is 465 years. The vertical
pressure scale height of the disk is $H_p=c_s/\Omega_K=6\,\mathrm{au}$.  We now
apply the perturbation analysis for wave numbers $k$ corresponding to
dimensionless wavelengths $\lambda=2\pi/k$ in the range between the smallest
possible wavelength $\lambda=H_p/r_0$ and the largest reasonable one
$\lambda=1$. The choice of $H_p/r_0$ is the smallest wavelength is based on
the assumption that the turbulent viscosity in the disk cannot lead to
radial structures that are narrower than about one vertical scale height. In
the current example this means that the smallest wavelength we should consider
is $\lambda=0.1$.

\begin{figure}
  \centerline{\includegraphics[width=0.52\textwidth]{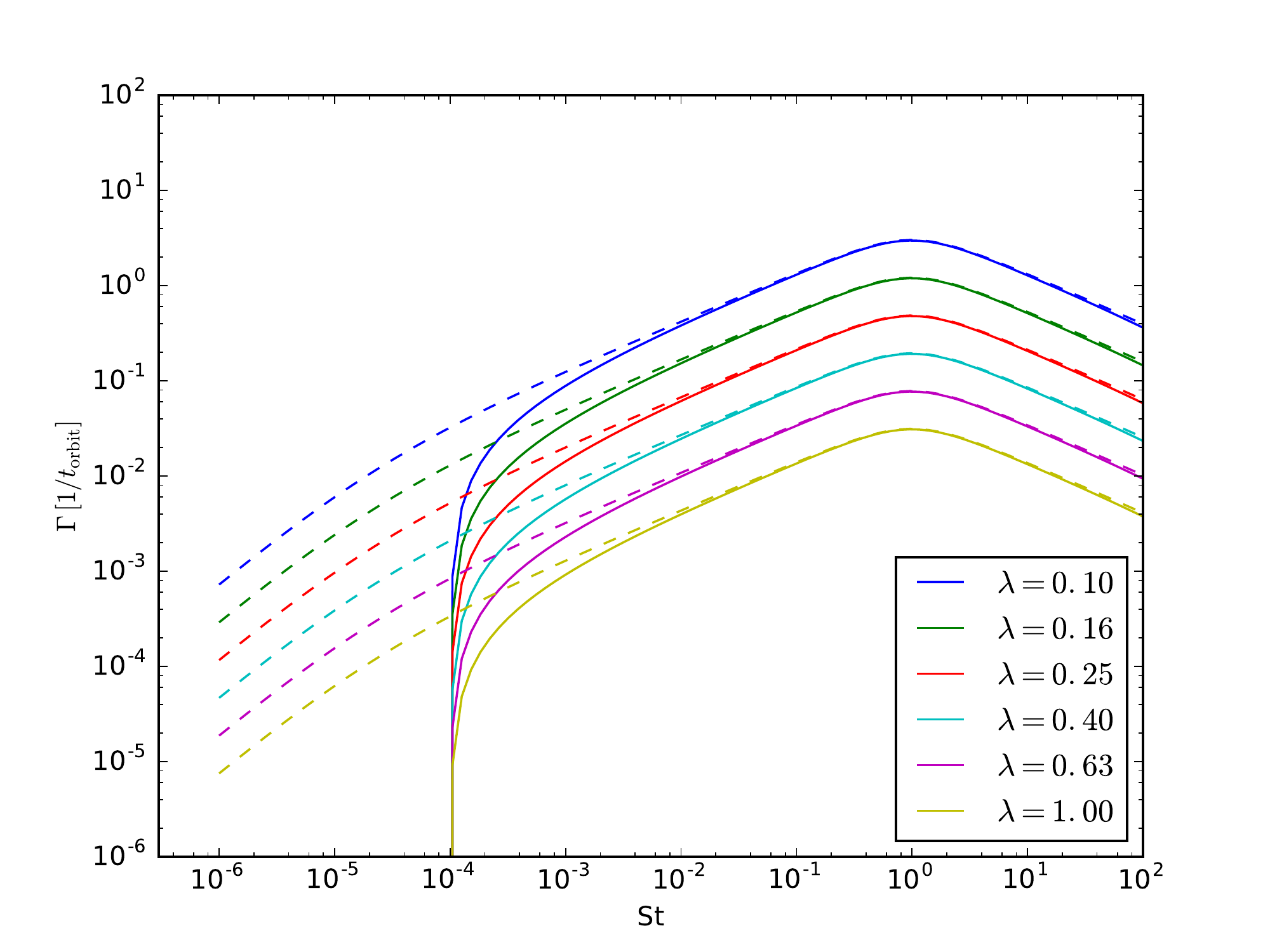}}
  \caption{\label{fig-gammaorb-simple}Growth rate in units of the reciprocal
    orbital time ($1/t_{\mathrm{orbit}}=\Omega_K/2\pi$) of the dust-driven
    viscous instability, as a function of dust particle size (expressed as
    Stokes number $\mathrm{St}$), according to the simplified analysis of
    Section \ref{sec-pert-analy-simple}. Dashed lines: case 1 (i.e.~$\phi_g=0$),
    solid lines: case 2 (i.e.~$\phi_g=-\phi_d$). The different lines show modes
    of different dimensionless wavelength $\lambda$ in the dimensionless
    coordinate $x$, where $\lambda=2\pi/k$. A dimensionless wavelength
    $\lambda=1$ means a wavelength as large as $r_0$. The parameters of the
    model shown here are $\phi_d=-1.0$, $\alpha_1=10^{-4}$,
    $M_{*}=1\,M_{\odot}$, $L_{*}=10\,L_{\odot}$, $r_0=60\,\mathrm{au}$,
    $T=43,\mathrm{K}$, $\mathrm{Sc}=1$ (i.e.~$c_s=0.39\,\mathrm{km/s}$ and
    $t_{\mathrm{orbit}}=465\,\mathrm{year}$).}
\end{figure}

We now compute the growth rate
\begin{equation}
\Gamma = \mathrm{Re}(i\omega)\comma
\end{equation}
of each of these modes for a range of different dust particle sizes. We express
the dust particle size in terms of its Stokes number $\mathrm{St}$ defined as
$\mathrm{St}=t_{\mathrm{stop}}/t_{\mathrm{orbit}}$, where $t_{\mathrm{stop}}$ is
the stopping time of the dust particle defined as
$t_{\mathrm{stop}}=f_{\mathrm{fric}}/m_{\mathrm{grain}}|{\bf
  v}_{\mathrm{grain}}-{\bf v}_{\mathrm{gas}}|$ where $f_{\mathrm{fric}}$ is the
friction force between the gas and the dust particle, $m_{\mathrm{grain}}$ is
the dust grain mass, and $|{\bf v}_{\mathrm{grain}}-{\bf v}_{\mathrm{gas}}|$ is
the absolute value of the velocity difference between the gas and the dust
particle. The precise translation between particle mass $m_{\mathrm{grain}}$ and
Stokes number is not trivial to express, because it depends on much detailed
physics, such as the porosity or fractility of the dust aggregate, its size
compared to the gas mean free path, the gas density and temperature etc.  For
typical disk parameters a Stokes number of unity at 60 AU would correspond to a
compact silicate dust particle of about a centimeter or a decimeter, and is
smaller for smaller particles. We refer to the literature for an in-depth
discussion of the relation between particle size and Stokes number \citep[see
  e.g.][]{2010A&A...513A..79B}. For our analysis only the Stokes number is
relevant.  The results of the present analysis are shown in
Fig.~(\ref{fig-gammaorb-simple}).

\begin{figure}
  \centerline{\includegraphics[width=0.52\textwidth]{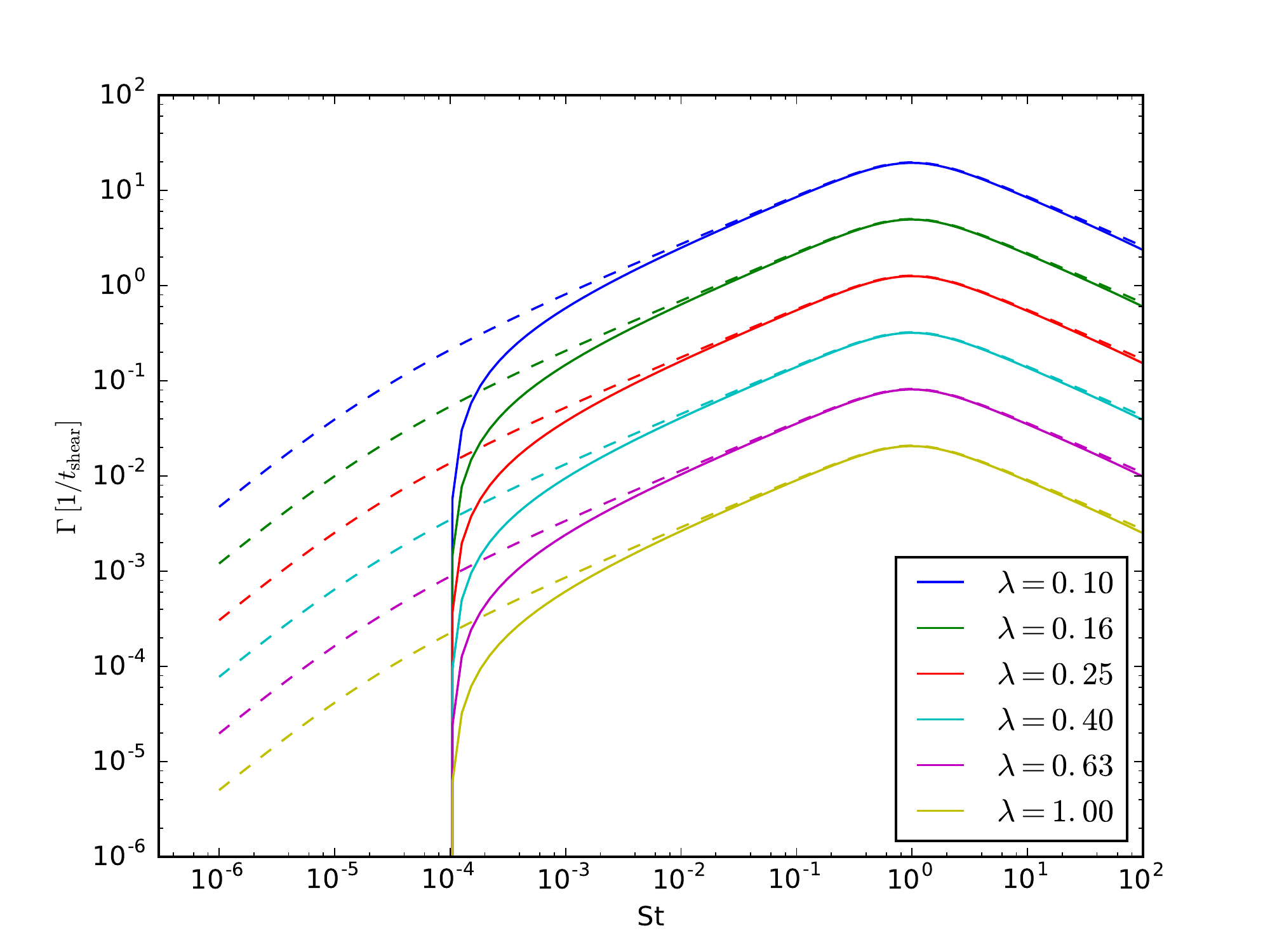}}
  \caption{\label{fig-gammalib-simple}Same as Fig.~\ref{fig-gammaorb-simple},
    but now with the growth rate $\Gamma$ expressed in units of the reciprocal
  shear time over a radial distance of $\Delta r=r_0\lambda$.}
\end{figure}

The growth rate $\Gamma=\mathrm{Re}(i\omega)$ is expressed in terms of the
reciprocal orbital time scale. This means that if this value is larger than
unity, the perturbation grows faster than the gas can orbit around the star.
This would not lead to `grand-design' rings, but instead to small arc-shaped
clumps, because if a perturbation is triggered at some azimuthal position, the
information about this event does not have time to propagate around the entire
orbit before the perturbation has grown to much larger amplitude. In other
words: to create global-scale rings we need a slow instability (a
`secular instability'). It must be slow compared to the time it takes for a
perturbation to shear out over $2\pi$ in azimuth. This time scale depends on the
radial width of the perturbation, which is related to the dimensionless
wavelength $\lambda$ of the unstable mode through $\Delta r=r_0\lambda$. Through
keplerian orbital dynamics we can then define this shear time scale
as
\begin{equation}\label{eq-time-shear}
t_{\mathrm{shear}} = \frac{2}{3}\frac{1}{\lambda}\,t_{\mathrm{orbit}}\fullstop
\end{equation}
So instead of comparing $\Gamma$ to the reciprocal orbital time, we should
express $\Gamma$ in terms of $1/t_{\mathrm{shear}}$. If we do so, the curves for
small $\lambda$ will move up. The results are shown in
Fig.~\ref{fig-gammalib-simple}. 
Whereever the curve lies above unity, the growth is faster than the azimuthal
communication. In that case small-scale arcs form instead of global scale
rings. Wherever the curve lies sufficiently below unity but above zero (which in
this log-representation means that the curve is visible in the plot), the
perturbation may lead to large scale rings. 

One can also see that, for case 2 ($\phi_g=-\phi_d$, solid lines in the figure)
the instability does not operate for $\mathrm{St}\le 10^{-4}$ (for this
set of model parameters), \revised{consistent with Eqs.~(\ref{eq-stability-criterion-simple-1},
  \ref{eq-stability-criterion-simple-2})}. This can be
understood because for very small grains (small $\mathrm{St}$) the dust is so
well-coupled to the gas that dust drift is virtually inhibited, meaning that
$\Sigma_d/\Sigma_g$ remains constant.

\begin{figure}
  \centerline{\includegraphics[width=0.52\textwidth]{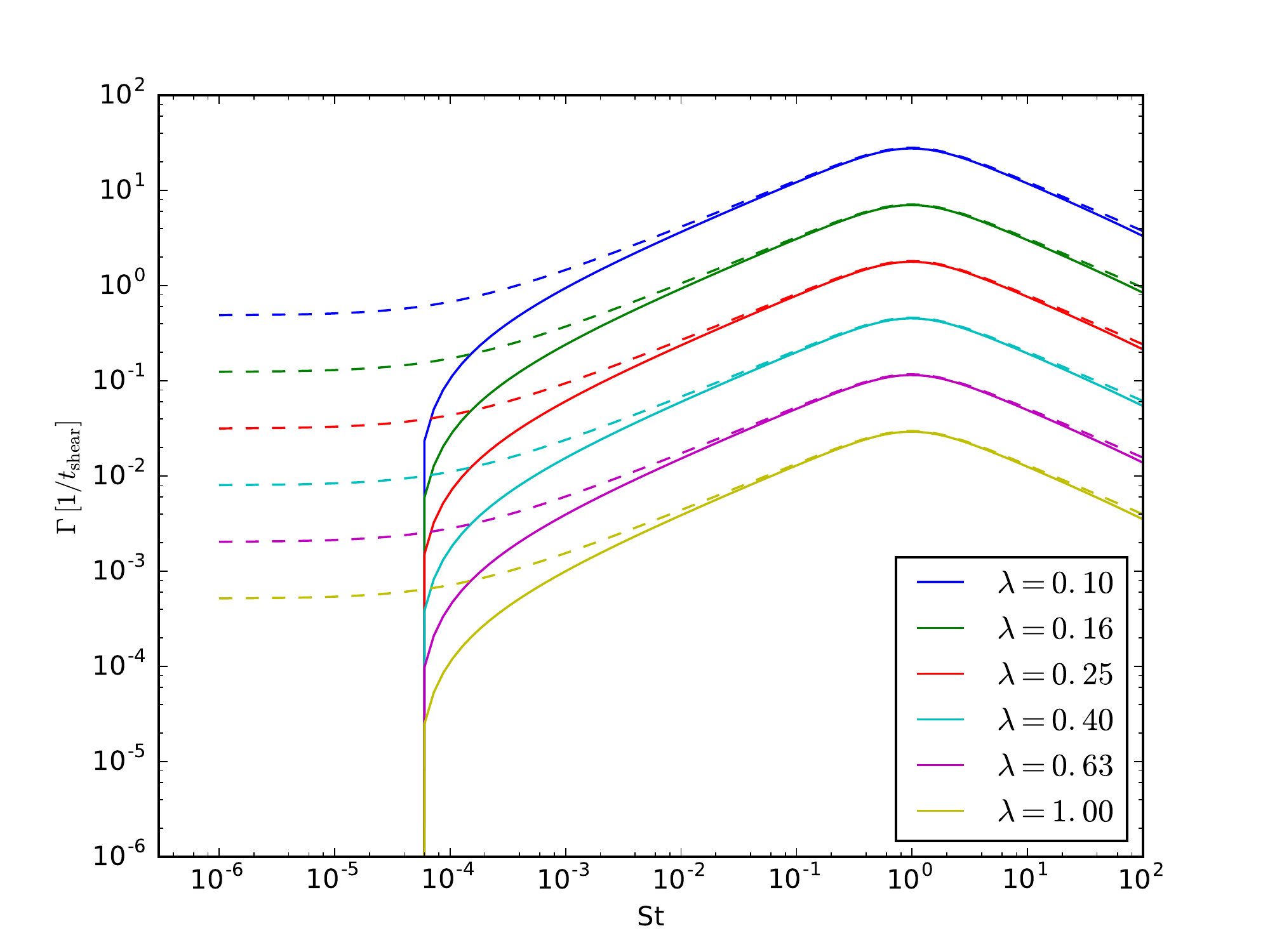}}
  \caption{\label{fig-gammalib-simple-phi2}Same as Fig.~\ref{fig-gammalib-simple},
    but now for $\phi_d=-2$ instead of $\phi_d=-1$.}
\end{figure}

\revised{For both case 1 and case 2, however, the instability does not occur for
  $\mathrm{St}\rightarrow 0$, i.e.\ for the case in which the dust does not
  drift at all. This changes if we set $\phi_d<-1$. In
  Fig.~\ref{fig-gammalib-simple-phi2} the results for $\phi_d=-2$ are shown
  (both case 1 and case 2). Now, at least for case 1 ($\phi_g=0$, dashed lines),
  the instability even operates for $\mathrm{St}\rightarrow 0$, i.e.\ without
  dust drift. The reason is that the convergent flow of gas, dragging along the
  dust with it, increases the gas and dust density enough to set the instability
  in motion. We then recover the instability by \citet{2015ApJ...815...99H}
  and others. We discuss this in Section \ref{sec-discussion}. 
}

As can be seen, however, the strongest growth occurs around Stokes numbers of
unity, and for the shortest wavelength $\lambda$. In this regime the growth rate
is faster than $1/t_{\mathrm{shear}}$. In a disk with a grain size distribution
spanning from tiny to large, this seems to suggest that the instability will be
mainly driven by the comparetively large $\mathrm{St}\simeq 1$ grains, which
would then lead not to rings but to numerous small arcs. Perhaps these arcs
can later merge into large scale rings is something that cannot be studied
using this linear perturbation analysis.

In reality the situation is likely more subtle. In a disk with a dust size
distribution it is typically the smallest grains that are affecting
the $\alpha$ the most, because the smallest grains have the largest total
surface area and can thus be most effective in removing free electrons and
ions from the gas. We therefore speculate that in spite of the strong
growth rate for $\mathrm{St}\simeq 1$ particles that results from our
analysis, it is mostly the smallest dust grains that drive the instability,
if at all. If most/all of the dust has Stokes numbers below the cut-off
for case 1, then if case 1 is applicable the instability would not operate
at all.

\subsection{A speculative scenario}
\label{sec-two-stage-scenario}
Let us speculate about the following scenario: We assume that only relatively
small dust affects the viscosity parameter $\alpha$. From
Fig.~\ref{fig-gammalib-simple} for, say, $\mathrm{St}\simeq 3\times 10^{-4}$ the
instability is driven at a low enough rate, even for the smallest wavelengths,
that a set of global rings can form. As these rings grow in strength, we will
enter into the non-linear regime. The radial derivative of the gas pressure will
start to display sign-changes, and thus form actual dust traps. Since the disk
also has large grains (even though they did not participate in the instability),
these large grains get trapped into the dust traps and form dense dust rings,
possibly even dominating the local density over the gas density.
\revisedtwo{At this point the frictional back-reaction of the dust onto the 
  gas will have to be taken into account, and the streaming instability
  \citep{2007ApJ...662..627J} may set in within these dust rings. Also}
the self-gravity of the population of large grains may start to play a role.
Perhaps a combination with the secular instability of
\citet{2016AJ....152..184T} could occur.  We are aware that these are mere
speculations, and more investigation (in particular: numerical modeling) is
required.

So far we have only looked at the growth rates of the modes, not their spatial
propagation. In other words, we looked at $\mathrm{Re}(i\omega)$ but not yet at
$\mathrm{Im}(i\omega)$. The above speculative scenario is only possible if the
initial ring-instability occurs more or less {\em in situ}, or in other words,
that it is not
a moving wave that is slowly amplifying but instead a standing wave that is
growing in amplitude. To verify this we need to study the ratio
$\mathrm{Im}(i\omega)/\mathrm{Re}(i\omega)$ for all cases where
$\mathrm{Re}(i\omega)>0$. For the simplified analysis of this section it
turns out that $\mathrm{Im}(i\omega)=0$. The mode grows exactly
in-situ, meaning that the above speculative scenario is plausible.

\section{Full perturbation analysis}
\label{sec-pert-analy-full}
The perturbation analysis for the full system of equations of Section
\ref{sec-basic-equations} is substantially more tedious than the simplified
analysis of Section \ref{sec-pert-analy-simple}, but the results are overall
consistent with each other. There are also some simplifications that we keep:
we still assume that the Stokes number does not change with time and space, and
the same holds for the Schmidt number. In reality, for a given particle size
the Stokes number changes if the gas density changes. Such effects are not
included.

\subsection{Stationary powerlaw solution}
\label{sec-stationary-solution}
Let us assume the following Ansatz for the stationary solution:
\begin{eqnarray}
\Sigma_{g1}(r) &=& \Sigma_{g0} \left(\frac{r}{r_0}\right)^p\comma\label{eq-ansatz-sigmag}\\
\Sigma_{d1}(r) &=& \Sigma_{d0} \left(\frac{r}{r_0}\right)^p\comma\label{eq-ansatz-sigmad}
\end{eqnarray}
where we deliberately took the same powerlaw index for both the dust and
the gas component. For the temperature profile we also assume a powerlaw
of the form
\begin{equation}\label{eq-ansatz-temp}
T(r) = T_0 \left(\frac{r}{r_0}\right)^q\fullstop
\end{equation}
The isothermal sound speed $c_s$ follows from this temperature by
Eq.~(\ref{eq-cs2-in-temp}).  The dimensionless vertical scale height $h$ is
defined as $h=H_p/r$, where the scale height $H_p$ is given by
Eq.~(\ref{eq-disk-hp}).

The radial gas velocity (Eq.~\ref{eq-vr-gas}) becomes
\begin{equation}\label{eq-vrg-as-dbllog-dir}
  v_{rg} = -\frac{3\nu}{r} \frac{\partial\ln(\Sigma_g\nu\sqrt{r})}{\partial \ln r}
  = -\frac{3\nu}{r}\left(p+\frac{1}{2}+\frac{\partial\ln\nu}{\partial \ln r}\right)\comma
\end{equation}
where $\nu$ is given by Eq.~(\ref{eq-def-nu}). It can be aposteriori verified
that stationary powerlaw solutions only exist if the $\alpha$ coefficient is
independent of radius, which we will, from here on, assume to be the case.
With Eq.~(\ref{eq-def-nu}) we then obtain
\begin{equation}\label{eq-vr-stationary_1}
  v_{rg} = -\frac{3\nu}{r}\left(p+q+2\right)\fullstop
\end{equation}
Inserting this into the gas continuity equation (Eq.~\ref{eq-gas-continuity}),
and setting the time-derivative to zero, yields $\Sigma_g\nu=$constant. This
means (with Eqs.~\ref{eq-def-nu}, \ref{eq-omega-kepler}, \ref{eq-cs2-in-temp},
\ref{eq-ansatz-temp}, \ref{eq-ansatz-sigmag}) that
\begin{equation}\label{eq-powerlaw-sum-steady}
p+q=-\frac{3}{2}\fullstop
\end{equation}
Inserting this into Eq.~(\ref{eq-vr-stationary_1}) yields
\begin{equation}\label{eq-vr-in-nu-r}
  v_{rg} = -\frac{3\nu}{2r}\fullstop
\end{equation}

Now let us do the dust, Eq.~(\ref{eq-dust-continuity}). Since by our Ansatz
$\Sigma_d(r)/\Sigma_g(r)$ is a constant (because both have the same powerlaw
index), the right-hand-side of Eq.~(\ref{eq-dust-continuity}) is zero. This then
immediately means that $v_d(r)/v_g(r)$ must also be a constant. If we now
look at the two terms in Eq.~(\ref{eq-vdust-1}), and we assume that $v_d(r)$
must have the same radial powerlaw dependency as $v_g(r)$, then both terms in
Eq.~(\ref{eq-vdust-1}) must have the same radial powerlaw dependency.  This
means that $c_sh$ must have the same radial powerlaw dependency on $r$ as
$v_{rg}$:
\begin{equation}
  \frac{d\ln(c_sh)}{d\ln r} = \frac{d\ln |v_{rg}|}{d\ln r} = q+\frac{1}{2}\comma
\end{equation}
where we used Eqs.~(\ref{eq-vr-in-nu-r}, \ref{eq-def-nu}) and the definition of
$q$ (Eq.~\ref{eq-ansatz-temp}) in the last
step. However, given that $c_sh=c_s^2/\Omega_Kr$ we already independently know that
\begin{equation}\label{eq-dlncsh}
  \frac{d\ln(c_sh)}{d\ln r} = q+\frac{1}{2}\comma
\end{equation}
which confirms that we indeed have a stationary powerlaw solution for both the
dust and the gas. We can now calculate the ratio of the dust radial velocity to
the gas radial velocity using Eq.~(\ref{eq-vdust-1}) with
Eq.~(\ref{eq-ansatz-sigmag}):
\begin{equation}\label{eq-vdust-vgas-stationary}
\begin{split}
  v_{rd} &= \frac{v_{rg}}{1+\mathrm{St}^2} 
  + \frac{c_sh}{\mathrm{St}+\mathrm{St}^{-1}}
  \left(p + \frac{q-3}{2}\right)\\
  &= \frac{1}{1+\mathrm{St}^2}\left[1+L\left(p+\frac{q-3}{2}\right)\right] \;v_{rg}\comma
\end{split}
\end{equation}
where we define
\begin{equation}\label{eq-definition-l}
L\equiv \frac{\mathrm{St}\,c_s^2}{\Omega_K\,r\,v_{rg}} = -\frac{2}{3}\frac{\mathrm{St}}{\alpha}\comma
\end{equation}
where in the last step we used the stationary solution for $v_{rg}$
(Eq.~\ref{eq-vr-in-nu-r}) and the equation for $\nu$ (Eq.~\ref{eq-def-nu}).

If we insert a standard example, $p=-1$, $q=-0.5$, then this becomes:
\begin{equation}
v_{rd} = \frac{1}{1+\mathrm{St}^2}\left[1+1.833\,\frac{\mathrm{St}}{\alpha}\right] \;v_{rg}\fullstop
\end{equation}

\subsection{Linearization}
\label{sec-linearization}
Now we impose a perturbation on the stationary solution. We introduce the
coordinate $x$:
\begin{equation}
r=r_0e^x\simeq r_0(1+x)\comma
\end{equation}
and the perturbations:
\begin{eqnarray}
\Sigma_g(r,t) &=& \Sigma_{g1}(r) (1+\sigma_g(x,t))\comma\label{eq-full-sigma-g-pert}\\
\Sigma_d(r,t) &=& \Sigma_{d1}(r) (1+\sigma_d(x,t))\comma\label{eq-full-sigma-d-pert}
\end{eqnarray}
similar to Section \ref{sec-pert-analy-simple}. Again, the subscript $1$ denotes
the stationary solution.  We keep the temperature and sound speed stationary
(i.e.~static in time, but varying in space). The perturbations $\sigma_d(x,t)$
and $\sigma_g(x,t)$ are allowed to affect $\alpha(x,t)$. We use the same recipe
for $\alpha$ as before (Eq.~\ref{eq-alpha-param}).  To first order in the
perturbations we can write:
\begin{eqnarray}
d\ln r            &=& dx\comma\label{eq-dlnr-dx}\\
d\ln\Sigma_g(r,t) &=& d\ln\Sigma_{g1}(r) + d\sigma_g(x,t) \comma\\
d\ln\Sigma_d(r,t) &=& d\ln\Sigma_{d1}(r) + d\sigma_d(x,t) \comma\\
d\ln\left(\frac{\Sigma_d(r,t)}{\Sigma_g(r,t)}\right) &=&
     d\sigma_d(x,t) - d\sigma_g(x,t)\comma\\
d\ln\alpha(r,t)   &=& \phi_d\,d\sigma_d(x,t)+\phi_g\,d\sigma_g(x,t)\comma\\
d\ln\nu(r,t)    &=& d\ln\nu_{1}(r) + \phi_d\,d\sigma_d(x,t) \nonumber\\
& & + \phi_g\,d\sigma_g(x,t)\fullstop\label{eq-differ-ln-nu}
\end{eqnarray}
Or specifically for the double-logarithmic derivative with respect
to $r$ we obtain (omitting the $(r,t)$ for notational convenience):
\begin{eqnarray}
\frac{\partial\ln\Sigma_g}{\partial\ln r} &=& p + \frac{\partial\sigma_g}{\partial x} \comma\label{eq-dbllog-sigma-g}\\
\frac{\partial\ln\Sigma_d}{\partial\ln r} &=& p + \frac{\partial \sigma_d}{\partial x} \comma\label{eq-dbllog-sigma-d} \\
\frac{\partial\ln(\Sigma_d/\Sigma_g)}{\partial\ln r} &=&
     \frac{\partial\sigma_d}{\partial x} - \frac{\partial\sigma_g}{\partial x}\comma\\
\frac{\partial\ln\alpha}{\partial\ln r} &=& \phi_d\,\frac{\partial\sigma_d}{\partial x}+\phi_g\,\frac{\partial\sigma_g}{\partial x}\comma\\
\frac{\partial\ln\nu}{\partial\ln r} &=& q+\frac{3}{2} + \phi_d\,\frac{\partial\sigma_d}{\partial x}+ \phi_g\,\frac{\partial\sigma_g}{\partial x}\fullstop\label{eq-diff-ln-nu}
\end{eqnarray}
The gas velocity $v_{rg}$ (Eq.~\ref{eq-vr-gas}) is now:
\begin{eqnarray}
  v_{rg} &=& -\frac{3\nu}{r} \frac{\partial\ln(\Sigma_g\nu\sqrt{r})}{\partial \ln r}\\
  &=& -\frac{3\nu}{2r}\left(1+2(1+\phi_g)\frac{\partial\sigma_g}{\partial x}+2\phi_d\frac{\partial\sigma_d}{\partial x}\right)
\comma\label{eq-vrgin-nu-with-perturb}
\end{eqnarray}
where we used Eqs.~(\ref{eq-diff-ln-nu}, \ref{eq-powerlaw-sum-steady}). The
dust velocity $v_{rd}$ (Eq.~\ref{eq-vdust-1}) becomes, using
Eqs.(\ref{eq-omega-kepler}, \ref{eq-cs2-in-temp}, \ref{eq-disk-hp},
\ref{eq-ansatz-sigmag}, \ref{eq-dbllog-sigma-g}) and the identities $h=H_p/r$
and $P=\rho_gc_s^2$:
\begin{eqnarray}
\label{eq-vdust-2}
  v_{rd} &=& \frac{v_{rg}}{1+\mathrm{St}^2} 
  + \frac{c_sh}{\mathrm{St}+\mathrm{St}^{-1}}
\left(\frac{\partial \ln \Sigma_g}{\partial \ln r} + \frac{q-3}{2}\right)\\
  &=& \frac{v_{rg}}{1+\mathrm{St}^2} \left[1+\frac{\mathrm{St}\,c_sh}{v_{rg}}
    \left(p+\frac{\partial \sigma_g}{\partial x} + \frac{q-3}{2}\right)\right]
\fullstop\label{eq-vrdin-nu-with-perturb}
\end{eqnarray}
To be able to use $v_{rg}$ and $v_{rd}$ in the viscous disk equations
Eqs.~(\ref{eq-gas-continuity}, \ref{eq-dust-continuity}), will be forced
to compute their radial derivatives, which is where the cumbersome math
comes in. To keep things as orderly as possible, we rewrite
Eqs.~(\ref{eq-gas-continuity}, \ref{eq-dust-continuity}) into double-logarithmic
form:
\begin{eqnarray}
\frac{\partial\ln \Sigma_{g}}{\partial t} +
  \frac{v_{rg}}{r}\frac{\partial\ln(r\Sigma_g |v_{rg}|)}{\partial\ln r}
  &=& 0\comma\label{eq-gas-continuity-doublog}\\
\frac{\partial\ln \Sigma_{d}}{\partial t} +
  \frac{v_{rd}}{r}\frac{\partial\ln(r\Sigma_d |v_{rd}|)}{\partial\ln r}
  &=& \nonumber\\
  \frac{1}{r\Sigma_d}\frac{\partial}{\partial r}
  \bigg(r{\cal D}_d & \Sigma_d & \frac{\partial}{\partial r}
  \ln\bigg(\frac{\Sigma_d}{\Sigma_g}\bigg)\bigg)\fullstop\label{eq-dust-continuity-doublog}
\end{eqnarray}
The double-logarithmic derivatives of $r\Sigma_{g/d}v_{r\,g/d}$ can, using
Eqs.~(\ref{eq-dbllog-sigma-g}, \ref{eq-dbllog-sigma-d}), be written out as
\begin{equation}\label{eq-dlog-cons-term}
  \frac{\partial\ln(r\Sigma_{g/d} |v_{rg/d}|)}{\partial\ln r}
  = 1 + p + \frac{\partial\sigma_{g/d}}{\partial x} +
  \frac{\partial\ln |v_{rg/d}|}{\partial\ln r}\fullstop
\end{equation}
The right-hand-side of Eq.~(\ref{eq-dust-continuity-doublog}) can also
be written into the derivatives of the perturbations. To first order in
$\sigma_{g}$ and $\sigma_{d}$ we get:
\begin{equation}\label{eq-simplified-rhs}
\begin{split}
  \frac{1}{r\Sigma_d}\frac{\partial}{\partial r}
  \bigg(r{\cal D}_d  \Sigma_d & \frac{\partial}{\partial r} 
  \ln\bigg(\frac{\Sigma_d}{\Sigma_g}\bigg)\bigg)\\
  &= \frac{{\cal D}_d}{r^2}
  \frac{\partial^2 (\sigma_d-\sigma_g)}{\partial x^2}\comma
\end{split}
\end{equation}
where we made use of the fact that $\partial(\sigma_d-\sigma_g)/\partial x$ is
already first order in $\sigma_{g}$ and $\sigma_{d}$, and that $\Sigma_d{\cal
  D}_d$ is, to first order in $\sigma_{g}$ and $\sigma_{d}$, constant. This is
because the stationary solution (Section \ref{sec-stationary-solution}) obeys
$\Sigma_{d1}{\cal D}_d\propto \Sigma_{g1}\nu$, which is constant.

What remains to be done is to derive expressions for the double-logarithmic
derivatives of the gas and dust velocities (Eqs.~\ref{eq-vrgin-nu-with-perturb}
and \ref{eq-vrdin-nu-with-perturb}, respectively) used in
Eq.~(\ref{eq-dlog-cons-term}). This is somewhat tedious algebra, which we defer
to Appendix \ref{app-dbllog-v}. After inserting the resulting expressions
(Eqs.~\ref{eq-dbllog-lin-der-vrg-2}, \ref{eq-dbllog-lin-der-vrd-2}), into
Eq.~(\ref{eq-dlog-cons-term}), we see that for both the gas and the dust version
the constant term $1+p$ drops out, and the expression of
Eq.~(\ref{eq-dlog-cons-term}) becomes linear in the perturbations. This
cancellation of the constant $1+p$ is not surprising, because it follows from
the fact that we start our perturbation analysis from the stationary solutions
of Section \ref{sec-stationary-solution}. Inserting the resulting gas- and
dust-versions of Eq.~(\ref{eq-dlog-cons-term}), together with
Eq.~(\ref{eq-simplified-rhs}), into the continuity equations
Eqs.~(\ref{eq-gas-continuity-doublog}, \ref{eq-dust-continuity-doublog}), we
find the following set of equations:
\begin{eqnarray}
  \frac{\partial\sigma_g}{\partial t}
  &+&\bigg[
    \tilde C_g\frac{\partial \sigma_g}{\partial x}
   + \tilde C_{gg}\frac{\partial^2\sigma_g}{\partial x^2}
   + \tilde C_{gd} \,\phi_d\left(
   \frac{1}{2}\frac{\partial\sigma_d}{\partial x} + \frac{\partial^2\sigma_d}{\partial x^2}\right)  \nonumber\\
   & & + \tilde C_{gd} \,\phi_g\left(
  \frac{1}{2}\frac{\partial\sigma_g}{\partial x} + \frac{\partial^2\sigma_g}{\partial x^2}\right)
  \bigg] = 0 \comma\label{eq-cont-gas-lin-2}\\
  \frac{\partial\sigma_d}{\partial t}
  &+&\bigg[
    \tilde C_d\frac{\partial \sigma_d}{\partial x}
   + \tilde C_{dg}\frac{\partial^2\sigma_g}{\partial x^2}
   + \tilde C_{dd} \,\phi_d\left(
  \frac{1}{2}\frac{\partial\sigma_d}{\partial x} + \frac{\partial^2\sigma_d}{\partial x^2}\right) \nonumber\\
& &    + \tilde C_{dd} \,\phi_g\left(
  \frac{1}{2}\frac{\partial\sigma_g}{\partial x} + \frac{\partial^2\sigma_g}{\partial x^2}\right)
  \bigg] =
 \tilde {\cal D}_d\,
    \frac{\partial^2(\sigma_d-\sigma_g)}{\partial x^2} \comma\label{eq-cont-dust-lin-2}
\end{eqnarray}
where the tilde-symbols are defined as:
\begin{eqnarray}
  \tilde C_g &=&\frac{v_{rg1}}{r}\comma\\
  \tilde C_d &=&\frac{v_{rd1}}{r}\comma\\
  \tilde C_{gg} &=& \frac{v_{rg1}}{r}C_{gg} \comma\\
  \tilde C_{gd} &=& \frac{v_{rg1}}{r}C_{gd} \comma\\
  \tilde C_{dg} &=& \frac{v_{rd1}}{r}C_{dg} \comma\\
  \tilde C_{dd} &=& \frac{v_{rd1}}{r}C_{dd} \comma\\
  \tilde {\cal D}_d &=& \frac{1}{r^2}{\cal D}_d\comma
\end{eqnarray}
where the symbols $C_{gg}$, $C_{gd}$, $C_{dg}$ and $C_{dd}$ are defined
in Eqs.~(\ref{eq-define-cgg}, \ref{eq-define-cgd}, \ref{eq-define-cdg},
\ref{eq-define-cdd}). We again insert trial functions
\begin{eqnarray}
\sigma_g &=& A e^{i\omega t-ikx}\comma\\
\sigma_d &=& B e^{i\omega t-ikx}\comma
\end{eqnarray}
and obtain the matrix equation
\begin{equation}
i\omega \left(\begin{matrix}
  A\\
  B
\end{matrix}\right)
=\left(\begin{matrix}
  M_{aa} & M_{ab}\\
  M_{ba} & M_{bb}
\end{matrix}\right)
\left(\begin{matrix}
  A\\
  B
\end{matrix}\right)\comma
\end{equation}
with
\begin{eqnarray}
  M_{aa} &=& ik\tilde C_g+k^2\tilde C_{gg} + \left(\tfrac{1}{2}ik+k^2\right)\tilde C_{gd}\phi_g\comma\\
  M_{ab} &=& \left(\tfrac{1}{2}ik+k^2\right)\tilde C_{gd}\phi_d \comma\\
  M_{ba} &=& k^2(\tilde C_{dg}+\tilde {\cal D}_d) + \left(\tfrac{1}{2}ik+k^2\right)\tilde C_{dd}\phi_g \comma\\
  M_{bb} &=& ik\tilde C_d+ \left(\tfrac{1}{2}ik+k^2\right)\tilde C_{dd}\phi_d 
  -k^2\tilde{\cal D}_d\fullstop
\end{eqnarray}
The eigenvalues, and thereby the growth rates of the modes, follow 
from Eqs.~(\ref{eq-gamma-pm}, \ref{eq-iomega-from-gamma}).

\subsection{Results}\label{sec-results-full}
\begin{figure}
  \centerline{\includegraphics[width=0.52\textwidth]{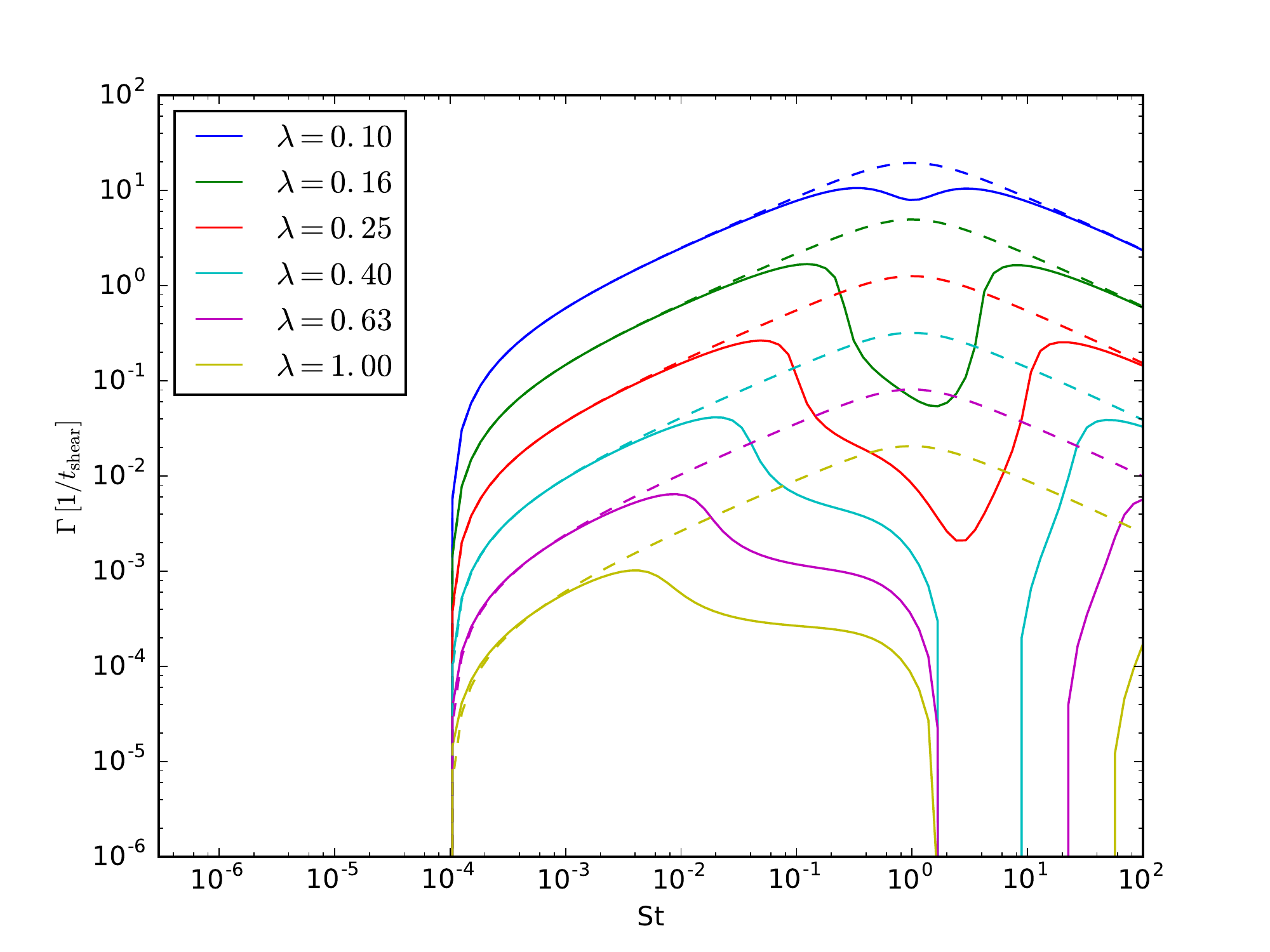}}
  \caption{\label{fig-gammalib-full-simple}Results for the full perturbation
    analysis of Section \ref{sec-pert-analy-full} (solid lines) compared to the
    results of the simplified perturbation analysis of Section
    \ref{sec-pert-analy-simple} (dashed lines). Here case 2 is shown
    ($\phi_g=-\phi_d$) with $\phi_d=-1$. For the rest the figure is the same as
    Fig.~\ref{fig-gammalib-simple}. \revised{If we would have plotted case 1
      instead of case 2, the difference would only be that the curves would not
      be cut off for $\mathrm{St}\lesssim 10^{-4}$, but would continue down to
      $\mathrm{St}\rightarrow 0$ in the same fashion as the dashed lines in
      Fig.~\ref{fig-gammalib-simple}.}}
\end{figure}

For the same parameters as in Section \ref{sec-results-simple} we
plot the resulting growth rates for the full perturbation analysis.
The result is shown in Fig.~\ref{fig-gammalib-full-simple} \revised{for case 2}.
It shows that for small enough particles and small enough wavelength $\lambda$
the simple perturbation analysis of Section \ref{sec-pert-analy-simple} agrees
well with the full perturbation analysis. \revised{The same is true if we would
  have plotted this diagram for case 1, the only difference to
  Fig.~\ref{fig-gammalib-full-simple} being, that the curves would continue down
  to $\mathrm{St}\rightarrow 0$ as in Fig.~\ref{fig-gammalib-simple}.} However,
for larger particles and/or larger wavelength, the full perturbation analysis
yields substantially weaker growth rates. But we see that for $\lambda\lesssim
0.25$ the growth rates are nevertheless everywhere positive where the simplified
analysis predicts positive growth rates. Since we need the instability to be
slow to obtain large scale rings, this is, in fact, advantageous for the
model.

\begin{figure}
  \centerline{\includegraphics[width=0.52\textwidth]{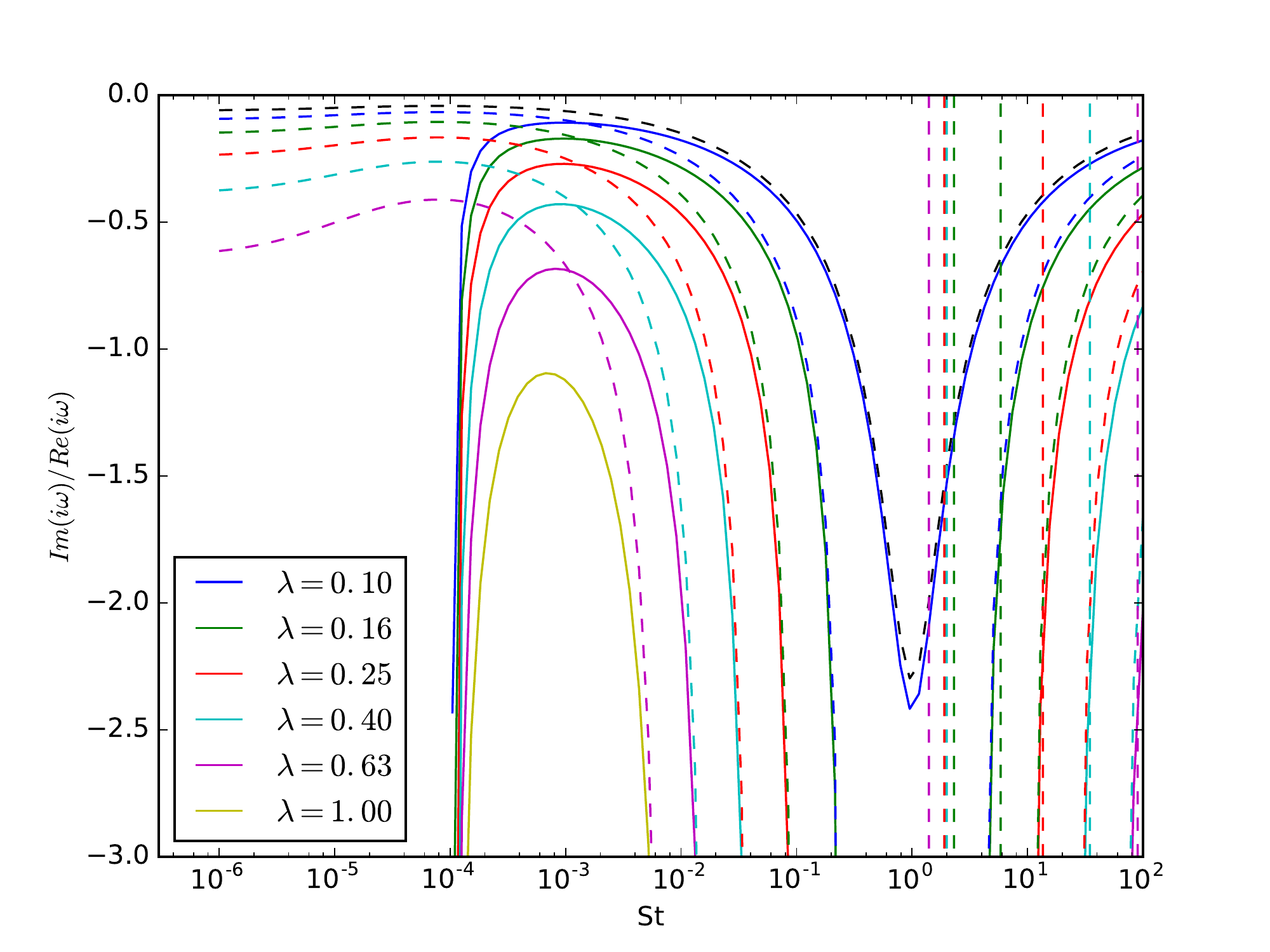}}
  \caption{\label{fig-imre-full}Ratio of
    $\mathrm{Im}(i\omega)/\mathrm{Re}(i\omega)$ for the full perturbation
    analysis. Dashed lines: case 1 ($\phi_d=-1$, $\phi_g=0$), solid lines: case
    2 ($\phi_d=-1$, $\phi_g=1$).}
\end{figure}

As we did for the simplified analysis of Section \ref{sec-two-stage-scenario} we
have to verify if the initial ring-instability occurs more or less in situ.  To
this end we plot $\mathrm{Im}(i\omega)/\mathrm{Re}(i\omega)$
(Fig.~\ref{fig-imre-full}). We see that, in contrast to the simplified analysis,
the imaginary component of $i\omega$ is not zero. However, we see in the plot
that for most Stokes numbers of interest $\mathrm{Im}(i\omega)$ is sufficiently
much smaller than $\mathrm{Re}(i\omega)$. That means that the growth can be
considered to be sufficiently well in-situ for the speculative scenario of
Section \ref{sec-two-stage-scenario} to remain plausible.

\section{Discussion}
\label{sec-discussion}
\subsection{Limitations and speculations}
The linear stability analysis of this paper shows that, at least in principle,
the combination of viscous disk theory, radial drift of dust, and the negative
feedback of the dust on the viscosity, could lead to ring-shaped patterns in a
protoplanetary disk.

The linear growth rate depends on the size of the dust grains, or more
precisely: on their Stokes number. Depending on the prescription of the
feedback, the instability is inhibited for the very small Stokes numbers.  But
for Stokes numbers beyond a critical value, the growth rate increases with
increasing $\mathrm{St}$. Around $\mathrm{St}\simeq 1$ the instability is
suppressed again. Fig.~\ref{fig-gammalib-full-simple} shows the growth rates as
a function of $\mathrm{St}$ for given wavelengths of the mode.

Not surprisingly, the instability grows the quickest for the shortest wavelengths.
The shortest wavelength is expected to be the disk pressure scale height, which is
therefore expected to be the dominant mode.

However, the feedback of the dust onto the viscosity of the gas is a
surface-area effect (dust grains removing free electrons and ions from the gas),
so one should expect the feedback to be the strongest for the smallest
grains. In our model we did not include this effect: we took the same feedback
recipe (Eq.~\ref{eq-alpha-param}) independent on grain size. We 
speculate that if a disk contains a size distribution of dust, the smallest
grains with Stokes numbers still beyond the critical one, will be the ones
that drive the instability. Since the growth rate of the instability for such
grain sizes is substantially slower than the time a blob would be sheared out
into a ring (see Fig.~\ref{fig-gammalib-full-simple}), the information about the
growth of the instability can be communicated over the full $2\pi$ azimuth of
the disk, so that a global ring is formed instead of a set of independent arcs.

With our model we cannot study what happens if the instability becomes
non-linear. Assuming we have a size distribution of dust grains, then, although
only the smaller grains drive the instability, also the larger grains will
undergo density enhancements: they will do so even stronger than the
instability-driving smaller grains. Once the mode becomes so strong that rings
of positive pressure gradient are produced, then the larger grains will
get trapped. Turbulent mixing always leaks a few of these grains
out of the traps, but on the whole, large grains would be trapped and produce
large-amplitude rings \revised{made from millimeter to centimeter size grains,}
similar to what is seen with ALMA in many sources. \revised{Our perturbation
  analysis suggests that the shortest wavelengths grow the quickest. However,
  viscous disk theory works on scales equal to or larger than the pressure
  scale height. The spacing between the rings in the gas structure of the disk will therefore 
  be at least a pressure scale height. The large dust grains, however, could
  conceivably get trapped in rings that are thinner than that.}

\revised{In this two-stage scenario (small grains triggering the rings, large
  grains getting trapped) the rings have to be sustained. If the small grains
  coagulate and become large, their ability to affect $\alpha$ reduces, and the
  rings may dissipate. Maybe this can be prevented through a bit of
  fragmentation of the pebbles, producing fresh fine grained dust. Even for low
  levels of turbulence the collision velocities of the pebbles may be relatively
  high. The typical turbulent eddy velocity at the top of the Kolmogorov cascade
  is $v_{\mathrm{eddy}}=\sqrt{\alpha}c_s$.  With a temperature of, say, 100 K
  one has $c_s=0.6\,\mathrm{km/s}$. If the fragmentation velocity is 1 m/s
  \citep[e.g.][]{2010A&A...513A..56G} one would need $\alpha\lesssim 3\times
  10^{-6}$ to prevent fragmentation. Anything above that would lead to the
  production of small grains. It is therefore feasible that a sufficient amount
  of small grains are continuously regenerated to keep the rings in place. On
  the other hand, aggregates made up of icy grains are thought to be more
  robust, and may fragment only at collision velocities of $\sim
  10\,\mathrm{m/s}$, or even up to $\sim 50\,\mathrm{m/s}$, depending on
  the size of the monomers of which they are made \citep{2009ApJ...702.1490W}}

\revised{Dust aggregates are most likely porous or fractal. Large `pebbles'
  may therefore still have rather large surface-to-mass ratios, and thus still
  strongly affect the ionization degree of the disk. We would, however, also
  observe them as if they were much smaller than they are, since also the
  optical properties of such dust aggregates depend a lot on the surface-to-mass
  ratio \citep{2014A&A...568A..42K}. The `big grains' that we identify as
  `big' due to their opacity slope at millimeter wavelengths therefore
  presumably also have a relatively low surface-to-mass ratio, and thus have
  little influence on the ionization degree of the disk. To have such big grains
  arranged in rings, we really need this two-stage scenario, as the big grains
cannot (according to our scenario) generate the rings themselves.}

\revised{There is, however, a big uncertainty with the model: the role of the vertical
structure. Small grains can be turbulently stirred to several pressure scale
heights above the midplane, even for relatively low turbulent $\alpha$. Is it
therefore still justified to assume that the wavelength of one pressure scale
height to be the strongest growing mode? Could it be that this would lead to
sufficient radial smearing that larger wavelength modes dominate?
\citet{2011ApJ...742...65O} study the effect of the vertical structure on the
viscosity of the disk. They conclude that the vertically averaged
magneto-turbulent viscosity only depends on the resistivity profile (and thereby
on the vertically averaged dust abundance) through three critical heights, and
is largely insensitive to the details of the resistivity profile itself. This may mean
that the vertically averaged dust abundance has, in most parts of the disk, only
limited influence (S.~Okuzumi, priv.comm.). }

Multi-wavelength
observations of the same sources can help answer these questions. For instance,
for TW Hydra both millimeter \citep{2016ApJ...820L..40A} and H-band observations
\citep{2017ApJ...837..132V} exist. \citet{2017ApJ...837..132V} study how the
rings in the millimeter compare to the large scale rings in the H-band. Some
correspondence is found, but overall the structures appear to be uncorrelated,
which appears to argue against our model, at least for this source.

\subsection{On rings and the slowness of the instability}
\revised{As argued in this paper, for an instability to lead to ring-like structures
rather than patchy/clumpy structures in a disk, the instability has to be slower
than the shear. This is the case for the dust-driven viscous instability discussed
in this paper. In hindsight this is not surprising, since the viscous time scale
of the disk can be quite long. One can quantify this by considering a perturbation with
radial width $W$ (i.e.~in our dimensionless form this is $W=\lambda r$). If
we express $W$ in units of the pressure scale height, which is the narrowest
viscous structures we can expect in the disk, we get}
\begin{equation}
W=w H_p = w\frac{c_s}{\Omega_K}\comma
\end{equation}
\revised{The viscous time scale for this perturbation is}
\begin{equation}
t_{\mathrm{visc}}=\frac{W^2}{\nu} = \frac{w^2}{\alpha\Omega_K}\fullstop
\end{equation}
\revised{The shear time (from Eq.~\ref{eq-time-shear}) is}
\begin{equation}
t_{\mathrm{shear}} = \frac{4\pi}{3}\frac{r}{wc_s}\fullstop
\end{equation}
\revised{The condition for the instability to be slow enough then becomes}
\begin{equation}
\frac{t_{\mathrm{visc}}}{t_{\mathrm{shear}}} = \frac{3}{4\pi}\frac{w^2}{\alpha}\frac{H_p}{r}\gg 1\fullstop
\end{equation}
\revised{The smallest possible value for $w$ is 1, which is also the fastest
  growing mode. This shows that if $\alpha$ is much smaller than the disk's
  dimensionless thickness (aspect ratio), then the instability (if it exists) is
  slow enough to create rings instead of patches. For a typical disk
  $H_p/r\simeq 0.05\cdots 0.1$ this means that $\alpha\ll 10^{-2}$ for the
  instability to be slow enough.  Therefore we can conclude that, if the rings
  seen in numerous disks are due to any form of viscous instability, the
  viscosity of the disk must be substantially lower than the canonical value of
  $10^{-2}$, or the instability must be slowed down by an even slower process
  such as dust drift of small enough dust grains.}

\subsection{Comparison to earlier work}
The ring-instability we have investigated in this paper appears to have a
relation to the ring-instability found by \citet{2005MNRAS.362..361W}. In their
model the disk had an active surface layer and a passive (`dead') midplane
layer. If gas would accumulate at some radius, the surface density of the active
layer stays the same, but that of the dead layer increases. In a vertically
averaged sense the viscosity thus gets reduced. This is mathematically identical
to our case of $\phi_d=-1$ and $\phi_g=0$ (prescription 1,
Eqs.~\ref{eq-alpha-param}, \ref{eq-case-1}), with $\mathrm{St}=0$. In our model,
however, we do not find an instability for these parameters. However,
\citet{2005MNRAS.362..361W} include the effect of the perturbation on the disk
midplane temperature, which we do not. In our case we indeed get the instability
if $\phi_d<-1$.

\revised{\citet{2015ApJ...815...99H} also studied the behavior of the
  two-layered disk model. They find that if the effective $\alpha$ of the
  two-layered disk is simply an average of the active and dead layers (weighted
  by their respective surface densities), then the disk remains stable. Our
  model with $\phi_d=-1$ and $\phi_g=0$ (prescription 1) and $\mathrm{St}=0$ (no
  dust drift) confirms this. When they apply a more sophisticated effective
  $\alpha$ recipe, based on \citet{2011ApJ...742...65O}, they find that the disk
  becomes unstable near the dead zone outer edge, because that is where the
  dependence of $\alpha$ on $\Sigma$ is the steepest. Our model confirms this,
  because the more sophisticated $\alpha$ recipe has a steeper dependence of
  $\alpha$ on $\Sigma$, which would amount, in our work, to $\phi_d<-1$ (again
  taking $\phi_g=0$ and $\mathrm{St}=0$), which we confirm to lead to
  instability. Our model is thus consistent with the earlier work by
  \citet{2015ApJ...815...99H}.  But by including dust drift our model is more
  general.}

\revised{\citet{2015A&A...574A..68F} perform 3-D full disk non-ideal MHD models
  and find ringlike structures, too, although rather wide ones and only two of
  them. But like \citet{2015ApJ...815...99H}, this is unrelated to dust drift.}

\revised{Our dust-drift induced viscous instability is, however, very similar to
  the instability found by \cite{2011IAUS..274...50J}. While the analysis in
  that paper is locally more detailed (including radial and azimuthal motions),
  our linear stability analysis includes the radial gradients and the global
  cylindrical geometry terms an is thus not just a local analysis. Furthermore
  our analysis includes the radial drift, as well as the turbulent diffusion,
  both of which play a key role in the mechanism. Moreover, we argue that the
  slowness of the instability is critical in getting grand-design rings
  rather than chaotic arc-shaped structures.}

\revised{The analysis in this paper} is
by no means a proof of the feasibility of this scenario. It will require
detailed 2-D/3-D viscous hydrodynamic disk modeling, and comparisons to
observations, to test this scenario.

\section{Conclusion}
\revised{In this paper we} show that it is conceivable that the combination
of viscous disk theory, radial drift of dust, and the negative feedback of the
dust on the viscosity, can produce (or at least trigger the formation of)
ring-shaped patterns in protoplanetary disks similar to what is seen in
protoplanetary disks at millimeter and optical/near-infrared wavelengths. 
\revised{From our present analysis we conclude:}
\begin{enumerate}
\item \revised{Even without dust drift, if $\partial \ln\alpha/\partial\ln\Sigma<-1$,
  the disk is prone to the viscous ring instability. This is a conclusion
  in agreement with work by \citet{2015ApJ...815...99H}.}
\item \revised{When dust grains are large enough to start drifting, yet small enough to
  have a substantial influence on the viscosity of the disk (through their
  ability to capture free electrons and ions from the gas), dust drift tends
  to cause a feedback loop on the disk viscosity, leading to dust-rich regions
  of low viscosity and dust-poor regions of high viscosity. In this way
  dust drift can trigger the viscous instability even when the disk would
  be stable otherwise ($\partial \ln\alpha/\partial\ln\Sigma\ge -1$). These
  findings are consistent with \cite{2011IAUS..274...50J}, and generalize them
  to global disk accretion with generalized viscosity description.}
\item \revised{The radial drift due to the global pressure gradient in the disk
  does not suppress the instability for small grains, but does so for grains
  with Stokes number near unity. The ring instability (viscous instability) must
  therefore be driven by small enough grains. This also agrees with the issue
  that small grains more easily affect the viscosity of the disk, because due to
  their larger surface-to-mass ratio, they more easily capture free electrons
and ions. }
\item \revised{For grand-design rings to form, such as those observed in real
  protoplanetary disks, rather than pseudo-random patchy structures, the
  growth rate of the instability must be slower than the shear-out time scale.
  Our analysis shows that this is generally the case, if the instability is
  driven by small enough grains and/or if the viscous $\alpha\ll 10^{-2}$.}
\item \revised{If the grains are too small, they hardly drift. Whether
  the disk is then stable or not depends strongly on the viscosity recipe.
  We have identified two cases: case 1 in which the absolute value of the
  dust density determines the viscous $\alpha$ and case 2 in which the 
    ratio of dust-to-gas density determines the viscous $\alpha$. For very
  small grains, the disk becomes stable for case 2, but may still become
  unstable for case 1, if the dependence of $\alpha$ on $1/\Sigma$ is steep
  enough.}
\item \revised{Observations seem to show that the rings seen at millimeter
  wavelengths are populated by relatively large rains. Since the viscous
  ring instability seems to be driven by small grains, this may seem inconsistent
  at first. We suggest that the initial rings are created by the viscous
  instability, which then produces strong enough dust traps that the
  larger grains get trapped and produce the ALMA images observed.}
\end{enumerate}

\begin{acknowledgements}
  This research was supported by the Munich Institute for Astro- and Particle
  Physics (MIAPP) of the DFG cluster of excellence ``Origin and Structure of the
  Universe''. We thank Satoshi Okuzumi for interesting discussions regarding the
  effect of the vertical structure. We thank the anonymous referee for insightful
  comments which helped improve the manuscript.
\end{acknowledgements}

\begingroup
\bibliographystyle{aa}
\bibliography{ms}
\endgroup

\appendix

\section{Double-logarithmic derivatives of the velocities}\label{app-dbllog-v}
The double-logarithmic derivative of the gas velocity with respect to $r$ is
computed from Eq.~(\ref{eq-vrgin-nu-with-perturb}) by initially taking the
single-logarithmic-derivative $\partial v_{rg}/\partial\ln r$, working out
$\partial(\nu/r)\partial\ln r$, and dividing again by
Eq.~(\ref{eq-vrgin-nu-with-perturb}), where one regularly makes use of the fact
that $|\sigma_{g/d}|\ll 1$ and that we expand only to first order in
$\sigma_{g/d}$.

For the double-logarithmic derivative of the dust velocity with respect to $r$
we start from Eq.~(\ref{eq-vrdin-nu-with-perturb}), and make use of the result
we already obtained for the double-logarithmic derivative of $v_{rg}$. Again we
regularly make use of the fact that $|\sigma_{g/d}|\ll 1$ and that we expand
only to first order in $\sigma_{g/d}$. To make the algebra more convenient
we define
\begin{eqnarray}
  L&=& \frac{\mathrm{St}\,c_sh}{v_{rg1}} = -\frac{2}{3}\frac{\mathrm{St}}{\alpha_0}\comma
\label{eq-defin-l}\\
  K&=& 1+L \left(p+ \frac{q-3}{2}\right)\comma\label{eq-defin-k}
\end{eqnarray}
where Eq.~(\ref{eq-defin-l}) is, in fact, the same as
Eq.~(\ref{eq-definition-l}).

After substantial algebra we find:
\begin{eqnarray}
  \frac{\partial\ln |v_{rg}|}{\partial\ln r} 
  &=&-(p+1) + C_{gg}\frac{\partial^2\sigma_g}{\partial x^2}\nonumber\\
  &&+C_{gd}\phi_d\left(
  \frac{1}{2}\frac{\partial\sigma_d}{\partial x}
  + \frac{\partial^2\sigma_d}{\partial x^2}\right)\nonumber\\
  &&+C_{gd}\phi_g\left(
  \frac{1}{2}\frac{\partial\sigma_g}{\partial x}
  + \frac{\partial^2\sigma_g}{\partial x^2}\right)\comma\label{eq-dbllog-lin-der-vrg-2}\\
  \frac{\partial\ln |v_{rd}|}{\partial\ln r} 
  &=&-(p+1) + C_{dg}\frac{\partial^2\sigma_g}{\partial x^2}\nonumber\\
  &&+C_{dd}\phi_d\left(
  \frac{1}{2}\frac{\partial\sigma_d}{\partial x}
  + \frac{\partial^2\sigma_d}{\partial x^2}\right)\nonumber\\
  &&+C_{dd}\phi_g\left(
  \frac{1}{2}\frac{\partial\sigma_g}{\partial x}
  + \frac{\partial^2\sigma_g}{\partial x^2}\right)\comma\label{eq-dbllog-lin-der-vrd-2}
\end{eqnarray}
with 
\begin{eqnarray}
  C_{gg} &=& 2\comma\label{eq-define-cgg}\\
  C_{gd} &=& 2\comma\label{eq-define-cgd}\\
  C_{dg} &=&2\left[1-\frac{L}{K}\left(p+\frac{q-4}{2}\right)\right] = \frac{12-4\,\mathrm{St}/\alpha_0}{6+11\,\mathrm{St}/\alpha_0}\comma\label{eq-define-cdg}\\
  C_{dd} &=&2\left[1-\frac{L}{K}\left(p+\frac{q-3}{2}\right)\right] = \frac{12}{6+11\,\mathrm{St}/\alpha_0}\label{eq-define-cdd}\fullstop
\end{eqnarray}
The second identities in Eqs.~(\ref{eq-define-cdg}, \ref{eq-define-cdd})
are for the case $p=-1$, $q=-1/2$.

\end{document}